\documentclass[amsmath,amssymb, amsthm, nofootinbib]{jpconf}
\usepackage{graphicx}

\def\demi{\frac{1}{2} }

\newcommand{\dr}{\rightarrow}

\def\cc{{\cal C}}
\def\dd{{\cal D}}

\def\hh{{\cal H}}

\def\lll{{\cal L}}

\def\nn{{\nonumber}}

\def\ss{{\cal S}}
\def\tt{{\cal T}}

\newcommand{\lalg}[1]{\mathfrak{#1}}
\newcommand{\SU}{\mathrm{SU}}

\renewcommand{\sl}{\lalg{sl}}

\def\ka{{\kappa}}
\newcommand{\mone}{^{-1}}

\def\tl{\widetilde}
\renewcommand{\v}{\overrightarrow}

\def\la{\langle}
\def\ra{\rangle}

\def\f{\frac}

\newcommand{\Ref}[1]{(\ref{#1})}

\newcommand{\schwa}{Schwarzschild }

\newcommand{\bem}{\begin{bmatrix}}
\newcommand{\eem}{\end{bmatrix}}
\newcommand{\beq}{\begin{equation}}
\newcommand{\eeq}{\end{equation}}
\newcommand{\beqa}{\begin{eqnarray}}
\newcommand{\eeqa}{\end{eqnarray}}
\newcommand{\arr}{\rightarrow}
\newcommand{\what}{\widehat}
\def\w{\wedge}
\def\be{ \begin{equation}}
\def\ee{\end{equation}}

\begin{document}
\title{Some comments on the universal constant in DSR}
\author{{\bf Florian Girelli}\footnote{girelli@sissa.it}}
\address{SISSA, Via Beirut 2-4, 34014 Trieste, Italy and INFN, Sezione di Trieste}

\author{{\bf Etera R. Livine}\footnote{etera.livine@ens-lyon.fr}}
\address{Laboratoire de Physique, ENS Lyon, CNRS UMR 5672, 46 All\'ee d'Italie, 69364 Lyon Cedex 07}


\begin{abstract}
Deformed Special Relativity is usually presented as a deformation of Special Relativity accommodating a new universal constant,
the Planck mass, while respecting the relativity principle. In order to avoid some fundamental problems (e.g. soccer ball
problem), we argue that we should switch point of view and consider instead the Newton constant $G$ as the universal constant.
\end{abstract}



\section*{Introduction}
There are hopes that  experiments (like GLAST, or LHC if there are extra dimensions) will soon measure   quantum gravitational
(QG) effects. Ideally, in order to derive precise predictions for the next experiments, we would like to integrate out all the QG
fluctuations (around the flat metric\footnote{assuming that the cosmological constant is zero.}) to obtain an effective action
for matter encoding  QG effects \be\label{action} \ss=\int d\phi_M dg\, e^{i\int \lll_M(\phi_M,g) + \lll_{GR}(g)} \,=\, \int
d\phi_M e^{i\int L_M(\phi)}, \ee where $g$ is the metric and $\phi_M$ represent all the matter and gauge fields other than
gravity. $\lll_M(\phi_M,g), \, \lll_{GR} (g)$ are respectively the lagrangian for matter and gravity.  The new effective
lagrangian $L_M(\phi_M)$ takes into account the QG fluctuations. We naturally expect non-trivial effects on the dynamics for
matter including a  mass renormalisation, higher derivatives, non-linearities and deformation of the energy conservation law.

This program can be carried out explicitly in three space-time dimensions in the spinfoam formalism. It was shown that the
effective dynamics for matter coupled to 3d QG is indeed a Deformed Special Relativity (DSR) \cite{el}. The four-dimensional case
is of course much harder, but heuristic arguments again point towards DSR (e.g \cite{DSRmotiv1, DSRmeasurement}). More generally,
DSR is usually formulated as implementing the Planck mass $M_P$ as a universal constant the same way that $c$ is a universal
speed in Special Relativity (SR) \cite{amelino}. However, in the same way that $c$ is a maximal speed in SR, $M_P$ becomes
naturally a maximal mass (or more generically an energy bound). This leads to confusion since the Planck mass can not be a
maximal bound for macroscopic systems. Even in formulations of DSR where $M_P$ is not a maximal bound, we run into the problem
that DSR effects grow with the mass of the system: a macroscopic system with a large mass will not behave classically as
expected. This is usually called the ``soccer ball" problem and is one of the main obstacles to the physical interpretation of
DSR.

We would like to argue here that a proper parallel between DSR and SR tells us that we should consider as the universal constant
not $M_P$ but the Newton constant $G$ instead (or more exactly $G/c^2=g$). This in particular leads us to consider as fundamental
a 5d space which is known to contain the DSR structure \cite{kowalski, GKKL}. We further argue along the lines that DSR is an
effective theory for matter coupled to gravity and show that some DSR features already emerge when taking into account that
particles locally deform the flat geometry into the \schwa metric.

\section{$G$ as a universal constant}

\paragraph{\textbf{Renormalisation and Choice of Units:}}
It is well known that coupling constants run when considering quantum corrections\footnote{$G$ could run also "classically" if
one believes in Modified Newtonian Dynamics: $G$ can be seen as dependent on the radial distance $r$ \cite{MOND}.} and QFT
renormalisation. This applies in particular to the Newton constant $G$. This can be explicitly seen when applying Renormalisation
Group technics to gravity \cite{reuter}: $G(k)$ is a function of the cutoff $k$. The latter is chosen according to the physical
situation.

In the search of phenomenology of QG, one would like to construct a theory encoding in an effective way the variations of $G$.
When considering the Newton's potential, we can always change point of view and consider a variable mass with a fixed $G$.
$$G(k)\frac{m}{r}\dr G\frac{m(k)}{r}. $$
In this sense we can take different points of view: either the mass is fixed with $G$ fluctuating or the mass is fluctuating with
$G$ constant. This can be interpreted as a choice of units\footnote{Peracci has emphasized that  the choice of units is important
when dealing with QG and can be related to a 5d picture \cite{percacci}.}. The  framework describing matter with  variable mass
has already been  studied by Bekenstein \cite{bekenstein}. To do physics, one needs to pick up a units system. One can first
choose a unit system which is independent of any particle data: the Planck units system, $M_P, \ L_P, \ T_P$.  If there is a
(spinless) particle present, one can also introduce different units systems depending on some  of the particle features. For
example if we consider a quantum particle the Compton (or de Broglie if massless) units  would be relevant, $ L_C, \
M_C=\frac{\hbar}{L_Cc}, \ T_C=c\mone L_C$.

First let us choose the Planck unit system. The fundamental constants $\hbar$, $ G$, $c$ are then in this units system
$c=L_PT_P\mone,$ $\hbar=M_PL_Pc, $ $ G=L_PM_P\mone c^{2}.$ Since the Planck units are fixed, the fundamental constants  do not
run in this unit system. We need also to express the characteristic of the particle in this unit system. The typical scale of the
particle is given by $L_C$ which is expressed in Planck unit $L_C=\chi L_P.$ Since $L_P$ is the minimum length we need to have
$\chi>1$. It is natural to have the same speed of light in the Planck unit system or the particle unit system\footnote{Note that
this could be relaxed, allowing then to have a variable speed of light \cite{joao}. The transformation for $G$ would then be also
more complicated.}, so that the time unit is just expressed as $ T=\chi T_P$. In this sense we are simply doing a conformal
transformation which can be local or global. Keeping $\hbar$ also fixed, the mass of the particle is then $m=\chi\mone M_P$. If
$\chi$ is just a global transformation, we are just doing a general rescaling. However if the transformation is local, we have a
particle with a varying mass $m$ with respect to the Planck units. $\chi$ can be also a more complicated function, for example
depend on the particle momentum. \be \label{variablem}m=\chi\mone M_P= \left(\alpha+ \sum_{n=1}\beta_n\frac{|\vec
p|^n}{M_P^{n}}\right)M_P, \textrm{ where } \alpha M_P=m_0, \textrm{ the rest mass}.\ee Conversely,  one can use  the Compton unit
system. The mass of the particle is then fixed since it is the unit. However if $\hbar$ and $c$ stay fixed in this choice of
units, the Newton constant $G$ becomes variable: $G'=\chi^3 L_P^3 \chi M_P\mone \chi^{-2}T_P^{-2}=\chi^2 G$. In the particle
units system, we have a fixed mass but a variable $G$. The idea is therefore to interpret  $\chi$ as encoding the QG fluctuations
in terms of the cutoff $k$, $G'=G(k)= \chi^2G$. If this cutoff is related to the particle momentum then we can expect to have
$\chi$ as in (\ref{variablem}) \cite{MDRRG}.

The Planck unit system seems to be more natural since it is independent on any matter content, it is  universal. In this unit
choice, $G$  is universal whereas the mass becomes a variable. We interpret this variability as encoding the QG fluctuations.

An important comment is now of order:  $M_P$ is the fundamental unit, however this does not mean that this should be
\textit{always} a maximum mass. Indeed for a quantum particle, $M_P$ can be seen as a maximum bound. On the contrary, it is easy
to see that if one considers a classical particle, $M_P$ is then a \textit{minimum mass}. In this sense, the Planck mass can be
seen as encoding the transition  from quantum  to classical. It surely can not be taken as a universal maximum mass for any body,
whereas the Newton constant $G$ is truly universal for any system in this choice of units.

\paragraph{\textbf{The Effective Field Theory Viewpoint:}}

We can further look at this issue from the effective field theory perspective. Computing loop corrections due to gravity in the
standard quantum field theory context, the $\hbar$ and $G$ factors usually combine to give a perturbative expansion in $E/E_P$.
The Planck energy/mass appears as a maximal mass scale if one is to make sense of such an expansion. In this context, a DSR
theory with a $M_P$ mass/energy bound looks completely natural and the conjecture is indeed that a certain resumming of some of
these  QG corrections can be interpreted as amplitudes of a DSR field theory with a $\kappa$-deformed Poincar\'e symmetry for a
deformation parameter $\kappa=M_P$.

We should however keep in mind that the apparent cut-off at $E_P$ is not a physical cut-off, but an artefact of the chosen
perturbative expansion. For energies larger than $E_P$, we expect non-perturbative effects. For instance, energies larger than
$E_P$ will not only induce Planck scale fluctuations around the flat metric but will deform the space-time on larger scales. We
would also need to take into account that the systems will actually start to gravitate. Nevertheless $E_P$ is not a physical
bound but the energy cut-off of the perturbative expansion. From the effective field theory point of view, one can change the
cut-off to a different value $\kappa\ge E_P$ by integrating out the relevant degrees of freedom. Applying this logic to DSR, we
could imagine starting from a DSR quantum field theory based at a given deformation parameter $\kappa$ (e.g the Planck mass).
Studying bound states with a mass larger than $\kappa$, we can hope that these bound states behave more and more classically as
their mass increases and are described by a DSR field theory with a renormalised deformation parameter $\kappa'$.  However,
extracting the effective physics of bound states of a quantum field theory is not an easy task.

This is the ``soccer ball" problem of DSR: showing that macroscopic systems behave classically despite the non-commutativity of
space-time at the microscopic scale. The parameter $\kappa$, usually set to the Planck mass $M_P$, should only be a cut-off,
which gets renormalised according to the considered physical situation. It should not be considered as a universal constant:
having allowed $\kappa$ to vary, the ``soccer ball" problem simply disappears. The universal constant is then the Newton constant
$g=G/c^2=L_P/M_P$. The issue then becomes how  to determine the cut-off $\kappa$ in term of the physical situation. This is
similar to determining the renormalisation scale $\mu$ in QFT depending on which experiment we do.

\paragraph{\textbf{DSR as a Semi-Classical Regime of QG:}}

In four space-time dimensions, $M_P$ and $L_P$ depend  explicitly on the Planck constant: $M_P\sim\sqrt{\f{\hbar}{G}}$,
$L_P\sim\sqrt{\hbar G}$. In the semi-classical limit defined by $\hbar\arr 0$, both Planck scales vanish. In particular, we do
not have a universal mass scale in that semi-classical regime\footnote{This is an essential difference between 3d quantum gravity
where one has $M_P\sim G\mone$, $L_P\sim\hbar G$.}. Therefore we can not expect DSR with a fixed universal mass scale
$\kappa=M_P$ to provide an effective description for quantum gravity in the semi-classical limit. DSR would instead correspond to
a regime where $\hbar$ stays of the same order than the Newton constant $G$, i.e a regime where gravitational effects and quantum
fluctuations are of the same magnitude. This behavior is expected at the Planck scale but not at the macroscopic scale. DSR with
fixed deformation parameter $\kappa=M_P$ seems fit to describe quantum gravity effects at a microscopic scale but obviously can
not apply to macroscopic objects in the semi-classical limit. The issue then becomes: which is the limit between microscopic and
macroscopic, i.e which physical objects/particles are described by DSR?

More precisely, as $\hbar$ varies and is sent to 0, keeping $G$ fixed, $M_P$ and $L_P$ do not remain fixed and finite but it is
their ratio $M_P/L_P=\frac{c^2}{G}$ does . Indeed $M_P$ and $L_P$ are both in $\sqrt{\hbar}$: as $\hbar$ varies, the mass scale
$M_P$ goes linearly the length scale $L_P$.
Thus we expect the semi-classical regime of quantum gravity to be described by a extended DSR theory with a variable parameter
$\kappa$ running linearly with the length scale. In this context, the key to the renormalization of the mass scale $\kappa$ is
that it depends on the Planck constant $\hbar$. This allows an interesting relation between  the DSR ``soccer ball" problem and
the quantum-classical transition. Indeed, for large composite systems, the effective Planck constant gets smaller as the system
grows bigger and therefore behaves more and more classically. In DSR, this leads to a renormalisation of $\kappa$.
This can be clearly seen when looking at the non-commutativity for a composite system: DSR predicts a non-commutativity of the
space-time coordinates controled by the mass scale $\kappa$. Denoting the space-time coordinates $X$ and the Lorentz generators
$J$, a generic commutation reads as:
$$
\left[\what{X},\what{X}\right]=\,i\hbar\left(a\f1\kappa \,\what{X}+ b\f\hbar{\kappa^2}\,\what{J}\right),
$$
where $a,b$ are arbitrary dimensionless constants. The important is that, when dealing with many copies of the same system, the
average space-time coordinates $\la X\ra_N\,\equiv\,(X^{(1)}+..+X^{(N)})/N$ do not have the same non-commutativity:
$$
\left[\la \what{X}\ra_N,\la \what{X}\ra_N\right]=\,i\hbar\left(a\f{1}{N\kappa} \,\la \what{X}\ra_N+
b\f{\hbar}{N^2\kappa^2}\,\what{J}\right).
$$
We therefore obtain a renormalization of the mass scale $\kappa\arr N\kappa$ with the size $N$ of the system. This is a crude
argument, assuming that the $N$ components do not strongly interact. Nevertheless, it shows that we do expect the
non-commutativity of space-time to get renormalized when considering composite systems: the fundamental non-commutativity, at the
Planck scale for instance, will be different from the effective non-commutativity felt by macroscopic objects.


\section{Extending Space-Time to a Space-Time-Mass}\label{ESR}
The great novelty of Einstein's Special Relativity is the concept of space-time, which marks the unification of time with the
three usual space dimensions. Indeed the speed of light $c$ is a universal constant defined for all observers
and allows the unification of time with space: an event is now described by a single 4-vector $x_\mu\,\equiv\,(ct,x_i)$. Then the
interpretation of $c$ as a maximal speed implies the existence of a light cone, which endows space-time with a non-degenerate
flat (pseudo-)metric: $v= \frac{dx}{dt}\leq c \Leftrightarrow c^2 dt^2- dx^2 \geq 0.$

Here, we would like to go one step further. We would like to argue that an effective theory for the kinematics and dynamics of
matter coupled to (quantum) gravity can be formulated in term of an extension of special relativity to a five-dimensional space,
where the fifth component can be identified as a function of the mass $\mathbf{m}$. In this sense the mass itself would become  a
variable and relative entity.


We assume the fundamental constant characterizing the gravitational interaction, the Newton constant $G$, to be universal. Using
dimensional analysis, $G$ allows to convert a mass $\mathbf{m}$ into a quantity with dimension of a length, $x_4=\frac{G}{c^2}\
\mathbf{m}= g \mathbf{m}$. We can then include the mass in a coordinate 5-vector generalizing the standard relativistic 4-vector:
\be x_A\,\equiv\,\left(ct,x_i,g\mathbf{m}\right). \ee We  expect this new five-dimensional Space-Time-Mass (STM) to become
relevant when  taking into account gravitation at an effective level in SR. Similarly to SR, we can further introduce a {\it mass
cone} structure, that is we put a maximum or minimum bound on $\frac{dx_4}{ds}$, where $s$ is the proper time. Thus we endow our
STM with the flat 5d metric $\eta_{AB}$ with signature $(+----)$: \beq\label{metricSTM1} dS^2\,\equiv\,c^2dt^2-dx_idx_i -g^2
d\mathbf{m}^2= ds^2-dx_4^2. \eeq Note that the $g$ is a function of $G$ and so is extremely small, explaining that we do not see
any mass fluctuations a priori. As an extension of SR, we propose to call this formalism Extended Special Relativity (ESR). Note
also that a similar construction has been proposed in \cite{wesson}.

At this stage the physical interpretation of this mass variable $\mathbf{m}$ still has  to be decided. A natural candidate is the
ADM mass $m$. In this case the mass cone can be interpreted as representing the \schwa ratio or the maximum energy ``density"
$\frac{m}{L}\leq\f{c^2}{2G}=\demi g\mone$. Choosing this interpretation leads to consider a physical object to be time-like with
respect in terms of the 5d metric.

On the other hand, a DSR particle is usually formulated as having a space-like trajectory in 5d space. Indeed it was shown in
\cite{GKKL} that the different formulations for the DSR particle can all be obtained as different gauge fixings of the same 5d
action: \be \label{action5d} S_{5d}=\int \, \pi^A {\rm d}y_A-\lambda(\pi^A\pi_A+\ka^2) -\mu(\pi_4-M), \ee where $\mu$ and
$\lambda$ are Lagrange multipliers and $M$ encodes the 4d rest mass. This looks in contradiction with the space-time-mass
introduced above. However, the DSR phase space coordinates $(y_A,\pi^A)$ are not straightforwardly related\footnote{ The
different DSR formulations give different prescriptions for the relation between the $(y,\pi)$ and the $(x,p)$. The Snyder basis
is the simplest one and defines the $x$ coordinates as a simple rescaling of the $y$'s \cite{GKKL}:
$$
p_\mu=\kappa\f{\pi_\mu}{\pi_4},\quad x_A=\f{\pi_4}\kappa y_A.
$$
Then the action reads $\pi^Ady_A=(p^\mu dx_\mu-\kappa dx_4)-(p^\mu x_\mu-\kappa x_4)d\ln\pi_4$ and the space-like condition
$\pi_A\pi^A<0$ simply means that the rest mass is bounded, $p_\mu p^\mu<\kappa^2$. Work on the precise relation with the
space-time-mass is still under investigation
 \cite{GLOR}.} to the standard space-time coordinates
$(x_\mu, p^\mu)$, so that we can not identify the newly introduced $x_A$ coordinates to the usual DSR coordinates $y_A$: the DSR
equivalent of $\mathbf{m}$ still has to be deciphered.

The 5d point of view allows a simple proposal to avoid the DSR problems such soccer ball problem, spectator problem or non
associativity as shown in \cite{equivDSR}. Indeed, since we work with a flat five-dimensional space with a theory invariant under
the 5d Poincar\'e symmetry, it is natural to consider the $\pi_A$ as the fundamental momenta and assume the the law of addition
of momenta $\pi$ to be trivial. The total momentum of a composite is then simply defined as the sum of the individual 5d momenta.
Considering two particles, the total momentum is then $\pi^{tot}=\pi^{(1)}+\pi^{(2)}$. Assuming that the two particles have the
same ``5d mass", $\kappa_1=\kappa_2=\kappa$, then $\kappa_{tot}$ will generically be different from $\kappa$: \be
\kappa_{tot}^2=-(\pi^{tot})^2=-(\pi^{(1)}+\pi^{(2)})^2=\kappa_1^2+\kappa_2^2-2\pi^{(1)}.\pi^{(2)}. \label{5daddition} \ee This is
the same as when dealing with the masses of particles in SR. This gives a precise implementation of the $\kappa$ renormalisation
for DSR in the framework of the STM. In this sense an Extended Special Relativity (ESR) based on the five-dimensional
space-time-mass and using a 5d momenta seems to be the correct framework to formulate consistently DSR.

\section{DSR phenomenology from gravitational effects}

The point of view that DSR is an effective theory for (quantum) gravity has been already advocated from many point of views (see
for instance \cite{el,DSRmotiv1,DSRmeasurement}). We would like to discuss here how simple gravitational effects naturally lead
to DSR-like structures.

In three space-time dimensions, gravity is a topological theory and particles are topological defects deforming space-time as
conical singularities. In the quantum gravity context, this feedback of matter on the geometry lead to a renormalisation of the
mass, a deformed non-commutative addition of momenta, a deformed Poincar\'e symmetry, and allowed to show that the effective
dynamics of particles (and more generally matter fields) is described by DSR \cite{el} (see as well appendix A).

In four space-time dimensions, gravity can be reformulated as an almost-topological field theory
\cite{l&a} and we could model particles as topological defects in a first order
approximation\footnote{Some of the features of 3d DSR might also apply to the 4d case. For example,
the momentum of the particle would again be defined through a non-local measurement on the
geometry: it would be measured as some kind of holonomy and would thus live in a curved manifold
(see e.g \cite{4dMM}). It might be too naive to consider the holonomy of the Lorentz connection on
a loop in space around the particle, since there is unique prescription to define a path ``around"
a point in a 3d manifold. It seems more promising to study observables associated to (spherical) 2d
surfaces around the particle/singularity, as also suggested in \cite{jerzy}. These observables
would correspond to some measurements of the curvature on these surfaces and we should be able to
reconstruct the particle's 4-momentum (or the corresponding energy-momentum tensor) from them. Here
we see an important difference with respect to the 3d case. Spheres of small radius around the
particle will feel a strong curvature and we are likely to obtain a strongly deformed addition of
the momentum, while spheres of large radius only weakly feel the presence of the particle and
everything should happen as in an almost flat space-time. In a sense, it is fairly natural to
expect a small $\kappa$ (i.e a strong deformation) when close to the particle but a very large
$\kappa$ (i.e a weak deformation) for length scales $L$ large compared to the Schwarzschild radius
$M\,G/c^2$ of the system. These considerations naturally lead again to a deformed addition of
momenta, to the non-commutativity of this addition and to the non-commutativity of the space-time
coordinates by duality.}. Here, we follow the simpler alternative to consider that the particle
deforms the flat space-time into the Schwarzschild metric. We will see that makes apparent some DSR
features and also explain the renormalisation of the $\kappa$ deformation parameter.

\paragraph{\textbf{Mass:}}
Gravity is an interaction and we need to take its energy into account for the definition of the mass of a system. More generally,
the mass of a system is not an easy concept to define in general relativity. We can for example consider the Brown-York formula
which computes the energy contained in a region of radius $R$ around the particle \cite{brown}. This takes into account the
self-energy of the particle: \be m=M(R)-g\f{M(R)^2}{2R} \,\Rightarrow\, M(R)=g\mone R\left(1-\sqrt{1-\f{R_S}{R}}\right), \ee
where $R_s=2gm$ is the Schwarzschild radius and $m$ the ADM mass (seen by an observer at infinity). This formula is valid from
$R=R_S$ to $\infty$. This renormalised mass decreases from $2m$ at $R=R_S$ down to $m$.

We see that the notion of mass becomes relative. Then which mass is the relevant one? Should we normalize the particle's momentum
to $m$ or $M(R)$ depending on the observer?

\paragraph{\textbf{Maximum mass:}}
We should also take into account the gravitational potential in the total rest energy of a many-particles system. For two
particles (at rest), the resulting total mass is then: \be \label{newton} M=m_1+m_2-g\f{m_1m_2}{r}, \ee where $r$ is the distance
between the two particles.  Keeping the length scale $r$ fixed, this deformed law of addition of masses leads to a maximal mass
$m_{max}= g\mone r$: we derived from simple principles the existence of a maximal mass scale $m_{max}\equiv\kappa$. Moreover this
mass bound $\kappa$ is not fixed but gets renormalised with the size of the system. Actually it scales linearly with the length
scale as expected in the semi-classical regime of DSR. Then if we fix the length scale to the Planck scale, $r=L_P$, we recover
as expected a mass bound $\kappa=M_P$.

We can try to make this argument more precise in the context of general relativity.  Then the mass bound of DSR-ESR is saturated
by the highest density state, i.e black holes. Looking at how to ``add" black holes with each other, we can deduce how the
maximal bound should get renormalized. Starting with two black holes, the Hawking area theorem states from the resulting total
horizon area is at least the sum of the two initial areas. Actually, accordingly to the holographic principle, the area  counts
the entropy of the black hole. Thus assuming a weak interaction and no additional degrees of freedom, the entropy will be
additive. Therefore considering than the horizon area scales as the mass squared $m^2$, the total mass resulting from the merging
of two black holes of mass $m$ will be $m_{tot}=m\sqrt{2}$. This is significatively less than the naive $2m$. The difference of
energy is actually due to the gravitational field which absorbs some energy during the process (and deforms the space-time). This
is generalizable to a system of $N$ particles of the same size. Assuming that their individual mass is bounded by the same mass
scale $\kappa$, the mass bound on the system of $N$ particles will be $\kappa^{(N)}=\kappa\sqrt{N}$ corresponding to $N$ black
holes of mass $\kappa$ merging into a single bigger black hole.

At the end of the day, we see that general relativity predicts a natural renormalisation of the mass bound $\kappa$ with the size
of the system. In a first approximation, this scaling goes as $\sqrt{N}$. From the 5d point of view, assuming the simple
additivity of the 5-momentum $\pi_A$ we see from equation \Ref{5daddition} that such a scaling corresponds in the case of a
two-particle system to the special configuration when the 5-momenta of the particles will be orthogonal to each other,
$\pi^{(1)}.\pi^{(2)}=0$.



\paragraph{\textbf{Modification of momenta addition:}}

Following the same logic as above, gravity will interact with the particles and contribute to the
energy/momentum of coupled systems. It will already modify the conserved quantities at the
classical level\footnote{Let us consider two Newtonian particles coupled by the gravitational
interaction:
$$
S=\int d\tau\,\left[\dot{\vec{x}_1}\vec{p}_1+\dot{\vec{x}_2}\vec{p}_2
-\f{\vec{p}_1^2}{2m_1}-\f{\vec{p}_2^2}{2m_2}+G\f{m_1m_2}{|\vec{x}_1-\vec{x}_2|}\right].
$$
The conserved energy is simply deformed by the gravitational potential: $E_{tot}=E_1+E_2-V$. On the
other hand, the total mass is still defined as $M=m_1+m_2$, the total conserved 3d momentum is
still $\vec{P}=\vec{p}_1+\vec{p}_2$ and the total angular momentum is conserved:
$$
\vec{J}=\left(\f{m_1\vec{x}_1+m_2\vec{x}_2}{m_1+m_2}\right)\w \vec{P}.
$$
As for the relative degree of freedom, we note $\vec{x}=\vec{x}_1-\vec{x}_2$ the relative position,
$\vec{p}$ the conjugate momentum and $m$ the reduced mass:
$$
\vec{p}=\f{m_2\vec{p}_1-m_1\vec{p}_2}{M}, \qquad m=\f{m_1m_2}{M}.
$$
The relative angular momentum $\vec{j}=\vec{x}\w \vec{p}$ is still conserved, but $\vec{p}$ is not
conserved anymore. Actually we need to introduce the new conserved Laplace-Runge-Lenz vector:
$$
\vec{A}=\vec{p}\w\vec{j}-m\f{Gm_1 m_2\vec{x}}{|\vec{x}|}.
$$
Finally, we can define a deformed relative momentum $\vec{q}\,\equiv\,\vec{A}\w\vec{j}/|j|^2$,
which is conserved but is not simply equal to the relative momentum. In terms of $x$ and $p$, the
new momentum reads:
$$
\vec{q}=\vec{p}-m\f{Gm_1m_2}{|x|}\f{\vec{p}-(\vec{p}.\hat{x})\hat{x}}{|p|^2-(\vec{p}.\hat{x})^2},
$$
where $\hat{x}\equiv\vec{x}/|x|$. }. Therefore, we naturally expect to obtain deformed addition for
the energy and momenta if one sticks to the notion of energy/momentum defined without interaction.


To actually derive precisely the addition of momentum, we would need to study the insertion of two
massive particles in a (flat) space-time i.e how two Schwarzschild metrics ``merge into" a single
metric. This is not an easy task since we do not know how to solve exactly the dynamics of two
black holes in general relativity. Thus we need to identify in which regime (strongly or weakly
coupled) we would like to work and develop an effective mechanics of black holes in general
relativity. More precisely, we should study the merging process in more details and derive the
total energy-momentum corresponding to a system of two black holes. It will certainly not be the
sum of the energy-momenta of the two black holes and we will obtain a deformation of the law of
addition of momenta generalizing the mass addition law which we discussed above. This would provide
us with an explicit proposal for a DSR-ESR addition law of momenta.

\paragraph{\textbf{5d formulation:}}

Finally, it would be interesting to be able to rigorously derive the fifth dimension from GR. A natural way to generate a fifth
dimension is to consider the renormalization flow of GR, the fifth dimension being the energy scale. As argued for example in
\cite{percacci}, it is very natural to endow this resulting 5d space with the AdS metric. Here, we show how  a fifth dimension
can emerge from a simple point of view.

Let us consider a particle in space-time and call its 4-momentum $p_\mu$, as hypothetically
measured by an observer at infinity (if the particle was the sole matter content of the
space-time). Then it is natural to assume that a real observer will actually measure a momentum
$\tl{p}_\mu=\alpha p_\mu$ where the coefficient $\alpha$ depends on the observer. $\alpha$ can come
from a non-trivial conformal factor or a mass renormalization or simply the slowing down of the
clock due to gravity\cite{GLOR}. In all cases, the parameter $\alpha$ reflects some dynamical
effect due to gravity. Thinking of $\alpha$ as a degree of freedom independent from the $p_\mu$'s,
it seems natural to introduce a 5-vector:
\begin{equation}\label{5dmomentum}
\pi_4=\alpha \kappa, \quad \pi_\mu= \tl{p}_\mu=\alpha p_\mu,
\end{equation}
where $\kappa$ is a mass scale introduced for the sole purpose to provide $\pi_4$ with the dimension of a moment. This natural
parametrization is similar to the Snyder's basis. Indeed, the 4-momentum is expressed in terms of the 5-momentum as
$p_\mu=\kappa\,\pi_\mu/\pi_4$. In fact on shell we can see that they definitely coincide.    Choosing the mass shell to be
$\pi_A\pi^A=\pm\kappa^2$, we obtain \be\alpha = \frac{\pm 1}{\frac{m^2}{\kappa^2}-1}.\ee According to the choice of sign of the
mass shell condition we have a maximum or minimum mass.

Now one needs to further identify the  gravitational degree(s) of freedom responsible for $\alpha$ and this will determine the
precise addition law for the 5-momentum $\pi_\mu$ and thus the deformed addition law for the 4-momentum $p_\mu$.

\section*{Conclusion}
Using different arguments (units, effective field theory, semi classical limit) we have showed that
to consider the  Newton constant (instead of the Planck mass) as the fundamental constant allows to
avoid the soccer ball problem plaguing DSR. Moreover, considering the \schwa ratio $G/c^2$ as
universal naturally leads to introduce a fifth dimension in configuration space related to the
notion of mass. Besides previous results showing that the 5d formalism for DSR allows to avoid
other issued e.g. the spectator problem \cite{equivDSR}, this provides us with yet another
motivation to work with DSR from the 5d perspective. We named this 5d approach ``Extended Special
Relativity" and explained how its phenomenology is related to gravity. The key question to answer
now is to understand the operational/physical meaning of this fifth coordinate in the DSR case.

\appendix \section{3d DSR from Quantum Gravity} Three-dimensional gravity can be quantized at the path integral level as a spin foam model, the
well-known Ponzano-Regge model (see \cite{pr1,pr2} and references therein). It was shown in \cite{el} that the spin foam
amplitudes for the coupled system gravity+particles define the Feynman diagrams evaluations of the DSR quantum field theory at
$\kappa=M_P$. The main input is that a particle\footnote{Here, we consider spinless particles. Spinning particle can be
understood in a similar way \cite{el}.} of mass $m$ simply creates a conical singularity in  space-time. Outside of the particle,
the space-time is still flat.

This deformation of geometry due to the feedback of  matter leads to corrections to the law of addition of momenta and energy
conservation.

Let us go into more details. In the following, we will restrict ourselves to the Euclidean 3d
space(-time) for simplicity, although the formulas can be adapted to the Lorentzian case by
considering the group $\SU(1,1)$ instead of $\SU(2)$. We also assume a trivial space-time topology.
The holonomy around a particle/singularity of mass $m$ is then an $\SU(2)$ group element, a
rotation of some given angle $\theta$:
$$
g=\cos\theta Id +i\sin\theta  \what{u}.\v{\sigma},
$$
where $\what{u}$ is a unit vector indicating the direction of the rotation and the $\v{\sigma}$ are
the three Pauli matrices. The deficit angle is $2\pi m/\kappa$, $\kappa=M_P$, so that the angle
$\theta$ is related to the mass by:
$$
\theta=2\pi\left(1-\f{m}{M_P}\right).
$$
The flat 3d momentum of the particle is defined as: \be g=\sqrt{1-\f{p^2}{\kappa^2}} Id +
i\f{\v{p}}{\kappa}.\v{\sigma}. \ee The first effect is a renormalisation of the mass:
$$
M^2\equiv \v{p}^2 =\kappa^2\sin^2\f{m}{\kappa}.
$$
The second effect is the deformation of the addition of momenta. Indeed, considering two particles,
the holonomy around the two particles will be the product of the two holonomies around each
particle:
$$
g_{tot}\equiv g_1 g_2.
$$
This implies that the total momentum is not the simple addition of the two momenta: \be \v{p}_{tot}=\v{p}_1\oplus \v{p}_2
=\sqrt{1-\f{p^2_2}{\kappa^2}}\v{p}_1+\sqrt{1-\f{p^2_1}{\kappa^2}}\v{p}_2-\f{1}{\kappa}\v{p}_1\wedge\v{p}_2. \ee The last term
involves a vector product and is therefore non-commutative. Its origin is the non-commutativity of the group product on $\SU(2)$.
Then the natural question is: should we take $g_1 g_2$ or $g_2 g_1$? This actually depends on the precise definition of the
holonomy. First, the holonomies are all defined with respect to a start point -- the observer. Then they are defined along a loop
from the start point around the particle. The holonomy is actually invariant under smooth deformations of the loop connected to
the identity. Let us assume that we have defined the holonomy $g_1$ around the particle 1. Then we have two choices for the
holonomy around the particle 2: we can go to the left of particle 1 and measure the holonomy $g_2^L$ or the right of the particle
1 and define $g_2^R$. The total holonomy around $1$ and $2$ is in the end uniquely defined as: $g=g_2^L g_1= g_1 g_2^R$.
Generically, $g_2^L$ and $g_2^R$ are different: they give different 3d momenta $\v{p}^L_2$ and $\v{p}^R_2$. We nevertheless know
the link between the two:  $g_2^R=g_1^{-1}g_2^Lg_1$ and so depends on the mass and momentum of the first particle. The message
here is that we must remember that we have defined the particle momentum   through a non-local measurement. This is crucial when
trying to interpret physically such DSR theories.

Finally, this deformed addition of momenta leads to corrections to the energy conservation law.
Indeed the condition $\v{p}_1\oplus \v{p}_2\oplus
\v{p}_3=0$ is different from $\v{p}_1+\v{p}_2+ \v{p}_3=0$ and contains correction terms in $\v{p}/\kappa$. These quantum gravity corrections are
interpreted as energy escaping to the gravitational field.

\medskip

Up to now, we have only described the momentum space of 3d DSR. We reconstruct the 3d space(-time) through the Fourier transform.
This Fourier transform is between functions on ${\bf R}^3$ and functions on $\SU(2)$: \be f(\v{x})=\int_{\SU(2)} dg\, e^{\f
i2{\rm Tr}\,g(\v{x}.\v{\sigma})}\, F(g), \ee where $dg$ is the Haar measure on $\SU(2)$ and we take the trace in the fundamental
two-dimensional representation. The plane wave mode actually reduces to the usual $\exp(i\v{x}.\v{p})$. The main difference is
contained in the Haar measure:
$$
dg=\f{d^3\v{p}}{\sqrt{1-\f{p^2}{\kappa^2}}}.
$$
The extra factor is simply $1/{\rm Tr} g$. The momentum cut-off at $\kappa$ leads naturally to a
minimal length, the Planck length $L_P\sim \hbar/\kappa$. Moreover, we define a star product
respecting the deformed law of addition of momentum:
$$
e^{ix.p_1}\star e^{ix.p_2}\equiv e^{ix.(p_1\oplus p_2)}.
$$
This product is dual to the convolution product on $\SU(2)$. And all the mathematical details on
these structures can be found in \cite{el}.

\medskip

The position operators are the translations on the momentum space $\SU(2)$ and are identified to the canonical $su(2)$ generators
$\v{J}$: $\v{X}=L_P\,\v{J}$. This leads to a deformed phase space:
\begin{eqnarray}
{[}X_i,X_j{]}&=&-2i\,L_P\epsilon_{ijk} X_k \nonumber\\
{[}X_i,p_j{]}&=&i\hbar\sqrt{1-\f{p^2}{\kappa^2}}\,\delta_{ij}-i\,L_P\epsilon_{ijk}p_k.
\end{eqnarray}
In the "classical" limit $\kappa\rightarrow\infty$, we recover the standard phase space symplectic
structure. Since the positions are non-commutative, we need to construct systems of
coherent/semi-classical states localizing space-time points. Such states are provided by the usual
$\SU(2)$ coherent states (see for example \cite{coherent}). They lead to an uncertainty in the
determination of the space(-time) coordinates of a particle depending on its distance from the
origin as $\delta X \sim \sqrt{L L_P}$. Such a square-root law is generally expected in DSR
theories \cite{coherent}.

\medskip

The last point of this review is that it is convenient to have a four-dimensional point of view on
this DSR theory. Indeed we can define the four-dimensional coordinates: \be \pi_\mu=\kappa\f 12
{\rm Tr}(g\sigma_\mu), \ee where $\sigma_0$ is the identity and the rest are the three Pauli
matrices. In details, we have:
$$
\pi_0=\kappa\sqrt{1-\f{p^2}{\kappa^2}},\quad \pi_i=p_i.
$$
Then $\SU(2)$ is defined as the sphere ${\cal S}^3=\{\pi_\mu\pi^\mu=\kappa^2\}$. Moreover the Haar measure simply reads: \be
dg=d^4\pi\, \delta(\pi_\mu\pi^\mu-\kappa^2). \ee it is therefore natural to write all the DSR expression and formulas and
scattering amplitudes in this 4d vocabulary. In fact, it is shown in \cite{ls} that the fourth dimension is necessary to define
an associative differential calculus and write down properly the DSR quantum field theory and its scattering amplitudes.


\end{document}